\documentclass{article}
\usepackage{times}
\usepackage{amsmath, amsthm, amssymb}
\usepackage[dvips]{graphics}
\usepackage{geometry}
\newcommand{\ket}[1]{|#1\rangle}
\newcommand{\braket}[2]{\langle #1|#2\rangle}

\newtheoremstyle{thm}
     {0pt}
     {0pt}
     {\itshape}
     {}
     {\bf}
     {.}
     {1em}
     {}

\theoremstyle{thm}
\newtheorem{thm}{Theorem}[section]

\newtheorem{cor}[thm]{Corollary}
\newtheorem{lem}[thm]{Lemma}
\theoremstyle{remark}

\theoremstyle{definition}

 \newtheoremstyle{note}
     {3pt}
     {3pt}
     {\itshape}
     {}
     {\bf}
     {:}
     {.5em}
     {}

  \theoremstyle{note}

\newenvironment{CompactEnumerate}{
 \vspace{-4pt}
 \begin{list}{\arabic{enumi}.}{%
     \usecounter{enumi} %
     \setlength{\leftmargin}{12pt}%
     \setlength{\itemsep}{0pt}
     }}
 {\end{list}}

\begin{document}
  \title{The Bayesian Learner is Optimal for Noisy Binary Search \linebreak (and Pretty Good for Quantum as Well)}
  \author{Michael Ben Or \thanks{The Hebrew University, Jerusalem, Israel}
  \and Avinatan Hassidim \thanks{The Hebrew University, Jerusalem, Israel}}
  \maketitle

\begin{abstract}



\noindent We use a Bayesian approach to optimally solve problems in
noisy binary search. We deal with two variants:

\begin{itemize}

\item Each comparison can be erroneous with some probability $1 - p$.

\item At each stage $k$ comparisons can be performed in parallel and
a noisy answer is returned

\end{itemize}

\noindent We present a (classic) algorithm which optimally solves
both variants together, up to an additive term of $O(\log \log(n))$,
and prove matching information theoretic lower bounds. We use the
algorithm to improve the results of Farhi et al \cite{FGGS99}
presenting a quantum (error free) search algorithm in an ordered
list of expected complexity less than $(\log_2n) / 3$.

\end{abstract}

\section{Introduction}\label{sec:intro}

%



Noisy binary search has been studied extensively (see
\cite{KMRSW80,Pel89,AD91,DGW92,BK93,FRPU94,Asl95, Mut96,Orr96,Ped99,
Pel02}). The basic model begins with an array of $n$ elements. We
are given a special element $s$, and try to find its rank in the
array. Every query consists of comparing $s$ to one of the elements.
One can add noise by making each comparison (or query) return the
wrong result with probability $1-p$. One can also think of an
adversarial model in which an adversary is allowed to choose whether
the algorithm gets the right answer. Our work focusses on the noisy
non-adversarial model.


Practical uses for optimal noisy search can occur (for example) in
biology. A simple application is eye tests, which can be considered
as comparing our sight capability to fixed benchmarks (determined by
the size of the letters we are trying to see). Other (more complex)
possible applications are trying to determine the supermolecular
organization of protein complexes and isolating active proteins in
their native form \cite{SCJ94,HEWJB04}. In both cases, the
3-dimensional conformation of the proteins should be conserved, and
solubilization methods are based on different percentages of mild
detergents. Further, the separation of the above molecules is based
on different percentages of acrylamide and bisacrylamide.
Determining the right percentage can be done by noisy binary search,
running a gel for each query.

An interesting theoretical use is another way to devise the results
of \cite{KK07}. They present a sophisticated algorithm to insert a
coin with an unknown bias to a list of coins with increasing bias
(which is also unknown). In order to use our algorithm, we need a
way to compare coins (an oracle). By using the clever reduction of
\cite{KK07} we can always assume that one of them is unbiased. We
can therefore flip both coins together, until we get different
results in both coins. We then assume that the coin that got heads
has higher bias towards heads, and consider this to be a noisy query
(the exact noise is dependent on parameters of the problem which
exist in \cite{KK07}).

Generalizing binary search (without noise) when k questions can be
asked in parallel and then answered together is trivial. The
algorithm is simply to divide the array into $k+1$ equal parts, and
ask in which of the parts is the element we are looking for. This
model, and its noisy variant, are important (for example) when one
can send a few queries in a single data packet, or when one can ask
the second query before getting an answer to the first.

\subsection{Previous Results}



It is known that one can search in $\Theta(\log(n)/I(p))$ queries
assuming probabilistic noise. One way of doing it is iterating every
query many times to obtain a constant error probability, and then to
travel the search tree backtracking when needed \cite{FRPU94}. This
leads to large constants, and has no easy generalization for the
batch learning model. Aslam showed a reduction of probabilistic
errors to an adversarial model (see \cite{Asl95,Ped99}), and stated
as an open question if it possible to achieve a tight algorithm.
Aslam's algorithm suffers from the same multiplicative factor that
arises in the adversarial algorithm, and might not be applicable to
generalizations of noisy search.


Although it is known that quantum binary search has complexity
$\Theta(\log(n))$, determining the exact constant remains an open
problem (\cite{FGGS99,JLB05,CLP06,BBHT98,Amb99,HNS02,CLP06}). Farhi
et al presented in \cite{FGGS99} two quantum algorithms for
searching an ordered list. They first presented a ``greedy''
algorithm with small error probability that clearly outperformed
classical algorithms. However, they could not analyze its asymptotic
complexity, and therefore did not use it. Instead, they devised
another algorithm, which can find the correct element in a sorted
list of length 52 in just 3 queries. Iterating this as a subroutine
gives an $0.53 \log_2 n$ quantum search algorithm. This was later
improved by \cite{CLP06} searching lists of 605 elements using 4
comparisons to get $0.433 \log_2 n$ queries. We note that these
algorithms are exact. Since Farhi et al's greedy algorithm has small
error probability iterating it on a fixed size list results in a
noisy binary search algorithm. However, without an exact analysis of
noisy binary search, the resulting bounds are not strong enough.




\subsection{Our Results}

The main intuition of our work  is simply to force the algorithm to
ask queries where it has no information about the answer, thus
causing it to be more exact. We do so by using a Bayesian learner
which tries to learn the place of the element we are looking for.
Note that in this case myopic behavior is optimal, although previous
(non optimal) algorithms were a lot more complex.

Assume that the element we are searching has equal probability to be
any element in the list. Partition the list so that both parts have
probability $1/2$ to contain the right element, and ask in which
part is our element by comparing it to the ``middle'' element (where
middle is being given by the probability measure). Following the
standard Bayesian approach update the probabilities of all elements
given the outcome. Iterate this (partitioning the array to ``equal''
parts, measuring and updating probabilities) until there are just a
few elements with relatively high probability to be the right
element, and then compare directly to these elements. In each
partition, we gain an expected $I(p)$ bits of information. Formally

\begin{thm} \label{main-classic}
There exists a (classic) algorithm which finds the right element in
a sorted list of $n$ elements with probability $1 - \delta$ using an
expected

\[
   \left\{ \begin{array}{l}
       \frac{\log(n)}{I(p)} + O(\frac{ \log \log
(n)\log(1/\delta)}{I(p)}) , \delta \le \log^3(n) \\
\frac{\log(n)}{I(p)} + O(\frac{ \log \log^2 (n)}{I(p)}) +
O(\frac{\log(1/\delta)}{I(p)}) , \delta \ge \log^3(n)
\end{array} \right.
\]

\noindent noisy queries, where each query gets the right answer with
probability $p$. This is tight up to log log terms.\end{thm}


We present a similar Bayesian strategy when we are allowed to use a
few queries in parallel (see \ref{gen-noisy-search}). Once we have
an exact noisy search algorithm, we can recursively use the noisy
greedy quantum binary search of Farhi et al. Measuring after $r$
queries in their algorithm corresponds to sampling the intervals
according to a probability distribution which is concentrated near
the correct interval. If the entropy of this distribution over the
$k$ equal probability intervals is $H_r$, then the average
information is $I_r = \log(k) - H_r$, and the expected number of
queries is $\frac{r \cdot \log(n)}{I_r}$. With this we can show

\begin{thm} \label{main-quantum} The expected quantum query complexity of searching an ordered
list is  less than $0.32 \log(n)$.
\end{thm}

We use our algorithms to prove some new quantum lower bounds on
noisy search, and on search which can have a probability of
failure.



Section 2 gives the classical algorithm, and proves the classical
lower bounds. Section 3 presents a quantum algorithm for searching
an ordered list. Section 4 improves the known lower bounds for
quantum binary search when the algorithm is allowed to err (even
with high probability).


\section{Classic Algorithm}\label{sec:new}

\subsection{Problem Settings}

Let $x_1 \ge \ldots \ge x_n$ be $n$ elements, and assume we have a
value $s$ such that $x_1 \ge s \ge x_n$, and we want to find $i$
such that $x_i \ge s \ge x_{i+1}$. The only way to compare $x_i$
and $s$ is by using the function $f(i) \rightarrow \{0,1\}$ which
returns 1 if $x_i \ge s$ and 0 if $x_i < s$. The problem is that
when calculating $f$ we have a probability of $1-p$ for error.
Note that calculating $f$ twice at the same place may return
different answers. As our approximation for $f$ has a chance of
error, we let our algorithm err with probability $\delta$. First,
we present an algorithm which is highly inefficient with respect
to $\delta$ but almost optimal (up to loglog factors) with respect
to $n$ and $p$, and then explain how to improve it.

The algorithm we present is based on using Bayes's formula to update
$\Pr(x_i \ge x \ge x_{i+1})$ for every $i$. To do that, we need a
prior for this distribution. To achieve a uniform initial
distribution, we apply a trick due to Farhi et al in \cite{FGGS99},
which doubles the initial search space, but turns the algorithm into
a translationally invariant one (thus making the prior uniform). The
idea is to add another element $x_{i+n}$ for each $x_i$, such that
all $2n$ elements are ordered in a circle. We
then apply the algorithm with a random shift on the circle, and thus
begin with a uniform prior.

Formally, Farhi et al. solve a different problem which is equivalent
to search. They define $n$ functions $f_{j}(x)$ defined by

\[ f_j(x) =
   \left\{ \begin{array}{l}
       -1, x < j \\ 1 , x \ge j \end{array} \right. \]

for $j \in \{1, \ldots, n\}$. A query in this problem is giving the
oracle a value $x$, and getting $f_j(x)$ for some fixed but unknown
$j$, and the goal of the algorithm is to find $j$. They then double
the domain of the functions and define $F_j(x)$ by

\[ F_j(x) =
   \left\{ \begin{array}{l}
       f_j(x),1 \le  x \le n \\ -f_j(x-n) , n + 1 \le x \le 2n \end{array} \right. \]

And use the fact that $F_{j+1}(x) = F_{j}(x-1)$ to analyze their
algorithm only for $j=1$. To do a similar trick, define $x_{n+1}
\ldots x_{2n}$ by $x_{i+n} = - x_{i}$. Note that if the algorithm
returns $r$ when given $f_r(x)$ as an oracle (remember that the algorithm does
not know that it queries $f_r$), it would return $r-k$ (mod $2n$) if
a shift $x_k$ would be applied to all its queries (that is whenever
the algorithm wishes to query a value $x$ it gets the value of
$f_r(x-x_k$) instead).

Before the algorithm begins, we choose a random shift $x_1 \ge x_k
\ge x_n$, and instead of calling $f_r(x)$ we use the oracle with
$f_r(x-x_k)$. This means that for any initial $j$ value such that
$x_j \ge s \ge x_{j+1}$, the probability that the right answer for
the modified algorithm is either $i$ or $i+n$ is $1/n$. This is true
because the new probability distribution is a convolution between
the old probability distribution (the value $j$) and the uniform one
(choosing $x_{k}$). We assume that this shift has been done and
return to our former definitions (i.e. $x_1 \ge \ldots \ge x_n$ with
the special element $s$ uniformly distributed).

{\bf Definitions} The algorithm uses an array of $n$ cells $a_1,
\ldots, a_n$, where $a_i$ denotes the probability that $x_i \ge s
\ge x_{i+1}$. The initialization of the array is $a_i = 1/n$, as we
have a flat prior distribution. Every step, the algorithm chooses an
index $i$ according to the values of $a_1, \ldots, a_n$, and queries
$f(i)$. After calling $f(i)$ the algorithm {\it updates} the
probabilities $a_i$. This means that if $f(i)$ returned 0 (i.e. $x_i
< s$ with probability $p$), we multiply $a_j$ for $j \le i$ by $p$,
multiply $a_j$ for $j > i$ by $1 - p$ and normalize so that the
values $a_1, \ldots, a_n$ sum up to 1. The exact action we take
depends on the sum $q = \mathop{\sum_{j=1}^i} a_j$. Assuming again
$f$ returned zero, the normalization is

\[ a_j =
   \left\{ \begin{array}{l}
       \frac{p a_j}{pq + (1-p)(1-q)}, j \le i \\ \frac{(1-p) a_j}{pq + (1-p)(1 - q)}, j > i \end{array} \right. \]

%

\noindent We write explicitly the update for $f(i)=1$

\[ a_j =
   \left\{ \begin{array}{l}
       \frac{(1-p) a_j}{(1-p)q + p(1-q)}, j \le i \\ \frac{p a_j}{(1-p)q + p(1 - q)}, j > i \end{array} \right. \]

\noindent Note that if $|p - 1/2| \gg |q - 1/2|$, as will be the
case in our algorithm, the normalization is almost multiplying the
probabilities by 2. For example, in the case $f(i) = 0$ we almost
have $a_j \rightarrow 2 p a_j$ for $j \le i$ and $a_j \rightarrow 2
(1-p) a_j$ for $j > i$.

\subsection{Algorithm}

The main idea of the algorithm is an intuitive generalization to
binary search. In every stage  partition the elements in the
"middle" and ask whether the middle element is smaller or larger
than $s$. The definition of "middle" depends on the probabilities of
the elements - we want to query an element $x_i$ such that $\Pr(x_i
\ge s) = 1/2$. There are two technicalities we must address:

\begin{enumerate}

\item It is not always possible to find an element such that
$\Pr(x_i \ge s) = 1/2$. Therefore, we use a constant called
$\epsilon_{par}$ ("par" stands for partition) which is an upper
bound to $|\mathop{\sum_{j=1}^i} a_j - 1/2| = |q - 1/2|$.
Its value will be chosen such that we are
optimal with respect to $p$. Enlarging this value will cause us to
extract less information each query.

\item It is hard to distinguish between elements which are very
close to each other. Therefore, the algorithm does not necessarily
finds the index of $s$, but rather an index $i$ such that there are
at most $l_{sur}$ elements between $x_i$ and $s$ ($l_{sur}$ stands
for surroundings). We can then iterate the algorithm, this time
searching the elements $x_{i - l_{sur}}, \ldots, x_{i + l_{sur}}$.
Making sure $l_{sur}$ is $O(\log(n))$ gives the right running time,
even if the constant in the $O$ notation is large (as this gives an
additive $O(\log \log(n))$ term to the runtime).
\end{enumerate}

The exact values for $\epsilon_{par}$ and $l_{sur}$ will be chosen
later.

\fbox{\begin{minipage}{15 cm}
\begin{CompactEnumerate}

\item \label{stop} If there is an index $i$ such that $a_i \ge
\epsilon_{par}$ we prove that $x_{(i - l_{sur})} \ge s \ge x_{(i +
l_{sur})}$ with probability greater than $1 - \delta/3$. It is now
possible to run recursively with $\delta \prime = \delta/3$ and
search in only $2l_{sur} + 1$ elements.

\item \label{part} Else find an index $i$ such that $1/2 -
\epsilon_{par} \le \mathop{\sum_{j=1}^i} a_j < 1/2$

\item Query $f(i)$ and {\it update} the probabilities. Return to
\ref{stop}.

\end{CompactEnumerate}
\end{minipage}}

Previous noisy search algorithms have already used weights, see for
example \cite{KMRSW80,BK93,KK07}. However, we choose weights
optimally, and use information even when $p$ is very small (see for
example the usage of $\epsilon_{good}$ in \cite{KK07}). This gives
us better results, and enables optimal generalization to the batch
model.

\begin{lem} If the algorithm reached stage \ref{part} it
is possible to find $i$ such that $1/2 - \epsilon_{par} \le
\mathop{\sum_{j=1}^i} a_j < 1/2$.
\end{lem}
\begin{proof}
Assume such $i$ does not exist. Let $k$ be the maximal value for
which $\mathop{\sum_{j=1}^k} a_j < 1/2$. This means that
$\mathop{\sum_{j=1}^{k+1}} a_j > 1/2$ and $\mathop{\sum_{j=1}^{k}}
a_j < 1/2 - \epsilon_{par}$, and therefore that $a_{k+1} >
\epsilon_{par}$, and we should have stopped in step \ref{stop}.
\end{proof}

We now need to prove two main claims - that we will end the
algorithm in step \ref{stop} in a reasonable time, and that when we
do so with high probability the value $s$ will be in the surroundings of
$i$. The first claim is stated as lemma \ref{algorithm-ends} and is
based on lemmas \ref{enough-infor} and \ref{inforamtion}. To address
state these lemmas we need to use the entropy $H(a_1, \ldots, a_n) = \mathop{\sum_{i=1}^n} - a_i \log(a_i)$
and the information  $I(a_1, \ldots a_n) = \log(n)
- H(a_1, \ldots a_n)$.


\begin{lem} \label{enough-infor} If $\forall i,$ $a_i < \epsilon_{par}$ then
$H(a_1, \ldots, a_n) \ge \log(1/\epsilon_{par})$.
\end{lem}
\begin{proof} $H(a_1, \ldots, a_n) = \mathop{\sum_{i=1}^n} - a_i \log(a_i)  \ge
\mathop{\sum_{i=1}^n} - a_i \log(\epsilon_{par}) =
\log(1/\epsilon_{par}) \mathop{\sum_{i=1}^n} a_i =
\log(1/\epsilon_{par})$

Where the first inequality comes from the monotonicity of the log
function and $\forall i,$ $a_i < \epsilon_{par}$.
\end{proof}

This means that if $H(a_1, \ldots, a_n) < \log(1/\epsilon_{par})$
There exists $i$ such that $a_i \ge \epsilon_{par}$


\begin{lem} \label{inforamtion}

In every iteration of the algorithm, the expected rise of the
information function $I(a_1, \ldots, a_n)$ is greater than $I(p) -
4 \epsilon_{par}^2(1-2p)^2 $ which is at least $I(p)(1 -
\frac{1}{3\log(n)})$ for $\epsilon_{par} = \sqrt{1/24log(n)}$.

\end{lem}

\begin{proof}

Let $b_1, \ldots, b_n$ be the new probability values (after we
update $a_1, \ldots , a_n$ according to the result of $f$). Assume
that the partition was between $k$ and $k+1$. Let
$\mathop{\sum_{i=1}^k}a_i = q$, and $N_{nor} = \frac{1}{pq +
(1-p)(1-q)}$ be the normalization constant used by the algorithm
in case $f(k)$ returned zero.  We look at the information for this
case:
$$I(b_1, \ldots, b_n |f(k)=0) = \log(n) +
\mathop{\sum_{i=1}^{k}}N_{nor}p \cdot a_i\log(N_{nor}p \cdot a_i)
+ \mathop{\sum_{i=k+1}^{n}}N_{nor}(1-p) \cdot a_i\log(N_{nor}(1-p)
\cdot a_i)$$ Where the $a_i$'s are the values before the {\it
update} and the $b_i$'s are the values after it. We analyze the
first sum

\begin{eqnarray*}  \mathop{\sum_{i=1}^{k}}N_{nor}p \cdot a_i\log(N_{nor}p \cdot a_i) =
& & N_{nor} p \log(N_{nor}p) \mathop{\sum_{i=1}^{k}}a_i + N_{nor}
p \mathop{\sum_{i=1}^{k}}a_i \log(a_i) =
\\& &  N_{nor} p  q \log(N_{nor}p) -  N_{nor} p H(a_1, \ldots, a_k)
\end{eqnarray*}

\begin{eqnarray*}
I(b_1, \ldots, b_n |f(k)=0) = & & \log(n) + N_{nor} p  q
\log(N_{nor}p) -  N_{nor} p H(a_1, \ldots, a_k) +
\\& &  N_{nor} (1-p) (1-q)
\log(N_{nor}(1-p)) -  N_{nor} (1-p) H(a_{k+1}, \ldots, a_n)
\end{eqnarray*}

To analyze the expected information gain, we look at the
probability for $f(k)=0$. Luckily, it is $pq + (1-p)(1-q)$, which
is $1/N_{nor}$. Calculating the information for $f(k)=1$ would
give similar results, but the normalization factor would change to
$M_{nor} = \frac{1}{p(1-q) + (1-p)q}$. The expected information
after the query is

$$I(b_1, \ldots, b_n |f(k)=0)/N_{nor} + I(b_1, \ldots, b_n
|f(k)=1)/M_{nor}$$

Looking on $I(b_1, \ldots, b_n |f(k)=0)/N_{nor}$ we can see that
\begin{eqnarray*}
I(b_1, \ldots, b_n |f(k)=0)/N_{nor} =
 & & \log(n)/N_{nor} +  p  q \log(N_{nor}) - q p \log(p) +  p H(a_1,
\ldots, a_k) +
 \\& & (1-p)  (1-q) \log(N_{nor}) + (1-q)(1-p)\log(1-p) -
(1-p) H(a_{k+1}, \ldots, a_n)
\end{eqnarray*}

Using $1/N_{nor} + 1/M_{nor} = qp + (1-p)(1-q) + p(1-q) + (1-p)q =
1$ we have
\begin{eqnarray*}
 I(b_1, \ldots, b_n |f(k)=0)/N_{nor} + I(b_1, \ldots, b_n
|f(k)=1)/M_{nor} =  \log(n) - H(p) - H(a_1, \ldots, a_n) + \\ pq
\log(N_{nor}) + (1-p)(1-q)\log(N_{nor}) + p(1-q)\log(M_{nor}) +
(1-p)q\log(M_{nor})
\end{eqnarray*}

\noindent Which means that the expected information increase after
the query is $pq \log(N_{nor}) + (1-p)  (1-q) \log(N_{nor}) +
p(1-q)\log(M_{nor}) + (1-p)q\log(M_{nor}) - H(p)$ Before we
simplify this further (and choose a value for $\epsilon_{par}$ to
make it close enough to $I(p)$) note that the expected increase
does not depend on the actual values of $a_1, \ldots, a_n$, or on
the information before the query (other than $q$).
\begin{eqnarray*}
pq \log(N_{nor}) + (1-p)  (1-q) \log(N_{nor}) +
p(1-q)\log(M_{nor}) + (1-p)q\log(M_{nor}) =
\\ (pq + (1-p)(1-q))\log(N_{nor}) + (p(1-q) + (1-p)q)\log(M_{nor})
=
\\- (1/N_{nor})\log(1/N_{nor})  - (1/M_{nor})\log(1/M_{nor}) =
H(1/N_{nor})
\end{eqnarray*}

We now need to bound $H(1/N_{nor})$. For an ideal partition $q =
1/2$ we will have $H(1/N_{nor}) = 1$, and the expected information
increase in each query would be $I(p)$, which is optimal. However,
$q$ deviates from $1/2$ by at most $\epsilon_{par}$, and we should
now choose $\epsilon_{par}$ small enough to get the desired
runtime. As $q \ge 1/2 - \epsilon_{par}$, we have

$$H(1/N_{nor}) \ge H(p + 1/2 + \epsilon_{par} - 2p(1/2 +
\epsilon_{par})) = H(1/2 + \epsilon_{par}(1-2p)) \ge 1 -
4\epsilon_{par}^2(1-2p)^2$$

Where the last inequality uses that if $1/2 \ge x \ge -1/2$ then
$1 - 2x^2 \ge H(1/2 + x) \ge 1 - 4x^2$

%

Manipulating this inequality gives $x^2 < \frac{1 - H(1/2
+x)}{2}$. Using this and substituting $\epsilon_{par} \le \sqrt{1
/ 24 \log(n)}$,

$$4\epsilon_{par}^2(1-2p)^2  = 16\epsilon_{par}^2(p - 1/2)^2 \le \frac{16(p-1/2)^2}{24\log(n)}
= \frac{2(p-1/2)^2}{3\log(n)} \le
\frac{1-H(p)}{3\log(n)}=I(p)/3\log(n)$$

Putting it all together, the expected information increase in
every stage is at least

$$H(1/2 + \epsilon_{par}(1-2p)) - H(p) \ge 1 - 4\epsilon_{par}^2(1-2p)^2 - H(p) \ge
I(p) - I(p)/3\log(n) = I(p)(1 - \frac{1}{3\log(n)})$$ which ends
the proof. \end{proof}

Note that $\epsilon_{par}$ is not a function of $p$.

\begin{lem}\label{algorithm-ends}
The algorithm will reach the recursion condition in stage
\ref{stop} in an expected number of $\log(n)/I(p) + O(1/I(p))$
function calls.
\end{lem}

\begin{proof}

By lemma \ref{enough-infor}, we need $H(a_1, \ldots, a_n) <
\log(1/\epsilon_{par})$. As the initial entropy is $\log(n)$ and
the expected information rise every stage is $I(p) (1- 1/3\log(n)
)$ (by lemma \ref{inforamtion}), we have that the expected number
of stages is at most

$$\frac{\log(n) - \log(1/\epsilon_{par})}{I(p)(1 - 1/3\log(n))} \le \frac{\log(n)}{I(p)(1 - 1/3\log(n))} \le \frac{\log(n)}{I(p)} +
2/3I(p)$$

Where we used $1/(c-x) < 1/c + 2x/c$ for $c > 2x \ge 0$.
\end{proof}

\begin{lem}\label{element-in} Suppose $a_i \ge \epsilon_{par}$ in step \ref{stop}.
Let $r = \frac{p(1-p)\log^2(1/\delta)}{2p-1}$, and $l_{sur} =
(\frac{p}{1-p})^{r} \frac{1}{\epsilon_{par}}$. Then with probability
$\ge 1 - \delta$ we have $a_{(i - l_{sur})} \ge s \ge a_{(i+ l_{sur})}$.

\end{lem}

\begin{proof}

As the lemma is symmetric we assume without losing generality that
$s > a_{(i - l_{sur})}$ and show that the probability for such a
distribution $a_{1}, \ldots, a_n$ is small. As the $a_j$'s sum up to
1, there is $k$ such that $i - l_{sur} \le k < i$ and $a_k <
1/l_{sur}$. This means that $a_i / a_k \ge \frac{\epsilon_{par}
p^{r}}{\epsilon_{par}(1-p)^r} = (\frac{p}{1-p})^{r}$. This ratio was
created by function calls $f(j)$ for elements $k < j < i$, such that
$f$ returned at least $x+r$ times 1, and at most $x$ times 0.
Considering the number of ones in $2x+r$ function calls in this
regime as a random variable, we get an expectancy of $(1-p)(2x+r) <
0.5(2x+r)$ and a  standard deviation of $\sqrt{p(1-p)(2x+r)}$. We
apply the Chernoff bound after making sure that for every value of
$x$ we have $x+r$ is at least greater than the expectancy by
$\log(1/\delta)$ standard deviations, or that

$$\mathop{min}_{x}  \frac{x+r - (1-p)(2x+r)}{\sqrt{p(1-p)(2x+r)}} \ge
\log{1/\delta}$$

Function analysis of this gives $x= \frac{r - p^r}{2p-1}$ and the
minimum is $\sqrt{\frac{r(2p-1)}{p(1-p)}}$. This gives $r =
\frac{p(1-p)\log(1/\delta)^2}{2p-1}$. Using the fact that for $1/2 <
p < 1$ and $a > 0$

$$(\frac{p}{1-p})^{a p (1-p)/(2p-1)} \le e^{a/2}$$

we get $l_{sur} < \log_2(e)/2\delta^2 \epsilon_{par} = O(1/
\delta^2 \epsilon_{par})$.
The dependency on $\delta$ can be improved
by another variant of the algorithm which will be described later.
\end{proof}

Lemma \ref{element-in} gives us the success probability of the
algorithm. Its expected runtime is the sum of two terms. By lemma
\ref{algorithm-ends} the expected runtime until $I(a_1, \ldots a_n )
> \log(n) - \log(1/ \epsilon_{par})$ is $\log(n)/I(p) + const/I(p)$.
By lemma \ref{element-in}, as $l_{sur} = O(\frac{1}{\epsilon_{par}
\delta^2}) = O(\frac{\sqrt{24 \log(n)}}{\delta^2})$ searching
between $i - l_{sur}$ and $i + l_{sur}$ adds another term of
$O(\log\log(n)/I(p))$ to the runtime.

{\bf Implementation Notes} We are interested in the query complexity
of the algorithm, rather than its runtime. However, we note that a
naive implementation of it is poly logarithmic in $n$ (actually
$O(\log(n)^2)$). This is done by uniting cells of the array $a_1, \ldots, a_n$
when there was no query which discriminates between them. We
begin the algorithm with a single segment which consists of the entire array.
Every query takes a segment, and turns it into two segments (so in the end
of the algorithm we are left with $O(\log(n))$ segments). After
each query the weight of each segment is updated ($O(\log(n))$ time)
and choosing where to ask the next query consists of going over
the segments (again $O(\log(n))$ time). This can be improved to $O(\log(n) \log \log(n))$
by saving the segments in a binary search tree. every edge on the tree
has a probability on it, such that multiplying the numbers
on a path between the root to a certain vertex gives the weight of
all the segments which are under the vertex (the leaves of the
tree each constitute of a single segment). Suppose we
need to query $x_j$, such that we already queried $x_k$, $x_l$,
$k < j < l$ and no other elements were queried between $x_k$ and $x_l$.
In this case the leaf which represents the segment $a_k, \ldots, a_l$
will have two sons, one representing $a_k, \ldots, a_j$ and the
other representing $a_{j+1}, \ldots, a_l$. According to the
result of the query, one son will have probability $p$,
and the other $1-p$. The data structure will then fix
the probabilities on the path between the root and the vertex
$a_k, \ldots, a_l$ according to the answer of the query.
Both finding the right element and updating the probabilities
takes time which is proportional to the depth of the tree.
Each query adds $1$ to the number of leaves, and therefore as
there are $O(\log(n))$ queries this will be the number of leaves.
Keeping the search tree balanced (such as by using Red and Black trees)
gives depth of $O(\log \log (n))$ as required.



\begin{thm} \label{classic-lower-bound}{\bf (Lower bound)} Let $A$ be a classical
algorithm which finds the right element in a sorted list, using
noisy comparisons. Assume that $A$'s success probability is $\ge
1- \tau$, then $A$ takes at least an expected
$\frac{\log(n)}{I(p)} - \frac{\log( 1 / (1 - \tau))}{ I(p)}$
comparisons. \end{thm}

\begin{proof}

We quantify the maximum amount of information gained every query.
Every oracle call gives us at most an expected $I(p)$ bits of
information. This means that after $\frac{\log(n)}{I(p)} -
\frac{\log( 1 / (1 - \tau))}{ I(p)}$ oracle queries, the algorithm
has $\log(n) - \log(1 / (1 - \tau))$ information bits. Knowing
where is the right element is $\log(n)$ bits of information. This
means that the algorithm has to guess at least $\log(1/( 1 -
\tau))$ bits of information, which is done with success
probability $1 - \tau$.
\end{proof}

\begin{cor} \label{classic-lower-bound-cor}{\bf (Lower bound without noise)} Let $A$ be a classical
algorithm which finds the right element with success probability
$\ge 1- \tau$, then $A$ takes at least an expected $\log(n) -
\log( 1 / (1 - \tau))$ comparisons. Moreover, with probability $1 - 2 \tau$ the algorithm uses
at least $\log(n) - 2\log( 1 / (1 - \tau))$ comparisons.\end{cor}

\subsection{Improving the Dependency on $\delta$}\label{dep-delta}
The problem with what we presented so far is the dependency on
$\delta$ in $l_{sur}$. Assume first $\delta < \log^{3}(n)$. Let
$l_{sur} = (1/\gamma^2)^{1/(2p-1)}$ for a constant $\gamma$. Keeping
the same halt condition, the probability to find the right element
when it is reached will be constant, and that with probability $1-
\delta$ we will find the right place for $s$ after $\log(1/\delta)$
trials. Note that this means that the algorithm will not end after
we are first stuck in stage 1. We therefore update the probabilities
of $a_1, \ldots, a_n$ even when we run the algorithm recursively. In
this variant the expected number of queries is $\frac{\log(n)}{I(p)}
+ O(\frac{\log(1/\delta) \log \log (n)}{I(p)})$. The dependency on
$\delta$ is what one would expect from this kind of algorithm. The
$\log \log(n)$ factor in the big-O notation comes from the recursive
part of the algorithm. Assume now $\delta > \log^{3}(n)$. Run the
algorithm with $\delta` = \log^{3}(n)$. After the algorithm
finishes, check $\frac{\log(1/\delta)}{I(p)}$ times if it returned
the right element. If the check succeeded, return this element. If
the check failed, start all over again, until the check succeeds.
The probability that the check fails is $1/\delta`$, and as $\delta`
= \log^{3}(n)$, the increase in the expected query complexity is
negligible.
This gives theorem \ref{main-classic}.

%
%
%
%
%
%

\subsection{Bounding the Variance of the Runtime}

So far we proved that our algorithm finds the right element with
probability $1- \delta$ with an {\em expected} number of
$\frac{\log(n)}{I(p)} + O(\frac{\log \log(n)}{I(p)\log(1/\delta)})$
queries. Using the strong lower bound in theorem
\ref{classic-lower-bound} we are able to bound the probability that
the number of queries needed is a lot greater than this number using
a generalized Markov inequality, which we do not prove:

\begin{lem} \label{generalized-markov} Let $X$ be a positive random variable such that $E(X) = a$.
Assume that $\Pr(X \ge b) \ge 1 - \beta$, then $\Pr(X > c) \le
\frac{a-b + \beta b}{c-b}$ for $c > a$.
\end{lem}

%
%
%
%
%

Assume that the expected number of queries needed is
$\frac{\log(n)}{I(p)} + \frac{c_1 \log \log(n)}{I(p)\log(1/\delta)}$
where $c_1$ is a constant.

\begin{lem}\label{bound}

Let $\chi > 1$ and $\delta > 0$. The algorithm presented before will
find the required element $s$ in an expected number of
$\frac{\log(n)}{I(p)} + O(\frac{\log \log(n)}{I(p)\log(1/\delta)})$
queries. The probability that the number of queries is greater than
$\frac{\log(n)}{I(p)} + \frac{\chi (c_1+2) \log\log(n)}{I(p)}$ is at
most $1/\chi$.

\end{lem}

\begin{proof}

We use the lower bound of theorem \ref{classic-lower-bound}, setting
$1 - \tau$ = $1 - 1/\log(n)$ (that is $\tau = 1/\log(n)$). According
to the theorem, this means that the number of queries is greater
than $\frac{\log(n)}{I(p)} - \frac{2\log(\log(n))}{I(p)}$ with
probability $1 - 2/\log(n)$. Using lemma \ref{generalized-markov},
with $a = \frac{\log(n)}{I(p)} + O(\frac{\log
\log(n)}{I(p)\log(1/\delta)})$, $b = \frac{\log(n)}{I(p)} -
\frac{2\log(\log(n))}{I(p)}$, $\beta = 2/\log(n)$ and $c =
\frac{\log(n)}{I(p)} + \frac{4\chi (c_1+2) \log\log(n)}{I(p)}$ we get
that the probability the algorithm requires more than
$\frac{\log(n)}{I(p)} + \frac{\chi 4(c_1+2) \log\log(n)}{I(p)}$
queries is smaller than $1/\chi$. \end{proof}


%

\subsection{Generalized Noisy Binary
Search}\label{gen-noisy-search}

In this section we generalize binary search. In the regular search,
the algorithm divides a sorted array of items into two parts, and
the oracle tells it in which part is the desired element. Our
generalization is to let the algorithm divide the sorted array into
$k+1$ parts, and the oracle will tell it in which part is the
correct element.

Generalizing the noise model, there is one right part and $k$ wrong
ones every query, so we need to state what would be the error
probability for each kind of mistake. This is done by adding $k+1$
probabilities (which sum up to 1), where the $h$'th probability
stands for the chance that the oracle would return $j+h \mbox{ (mod
k+1)}$ instead of the $j$'th interval\footnote{We could have
actually used $(k+1)^2$ numbers, stating the chance to get interval
$i$ instead of $j$ for all $i,j$. This would change the algorithm in
an obvious manner, and is not necessary for the quantum result.}.

Formally, let $g : \{1, \ldots, n-1\}^k \rightarrow \{0, \ldots,
k\}$. If $g$ is being given $k$ indexes, $i_1 > i_2 > \ldots > i_k$
it outputs the answer $j$ if $x_{i_j} \ge s \ge x_{i_{j+1}}$ when we
identify $i_0 = 0$ and $i_{k+1} = n$. The error probability is taken
to account by associating $k+1$ known numbers $p_0, \ldots, p_k$ to
$g$, such that if $x_j \ge s \ge x_{j+1}$ then the result $j+h \mod
(k+1)$ would appear with probability $p_h$.

The optimal algorithm for this case is very similar to the case $k =
1$ (which is $f$). In every step divide the array to $k+1$ parts
with (an almost) equal probability, and ask in which part is the
element we're looking for. Let $a_1, \ldots , a_n$, $\epsilon_{par}$
and $l_{sur}$ as before (albeit with different values this time).


\fbox{\begin{minipage}{15 cm}
\begin{CompactEnumerate}

\item If there is a value $i$ such that $a_i > \epsilon_{par}$
halt. If the algorithm halts, then with probability $1 - \delta/3$,
$x_{(i - l_{sur})} \ge s \ge  x_{(i + l_{sur})}$, continue recursively.

\item Else, let $i_1, \ldots i_k$ be indices such that the sum of
the elements between two indices does not deviate from $1/k$ by
more than $\epsilon_{par}$:

$$1/k - \epsilon_{par} \le \mathop{\sum_{h=i_{j-1}}^{i_j}}a_h  \le 1/k
+ \epsilon_{par}$$

\item Use $g(i_1, \ldots, i_k)$ and update the probabilities
according to Bayes's rule.

\end{CompactEnumerate}
\end{minipage}}

We use $\epsilon_{par} = \frac{1}{k} \sqrt{1/24 \log(n)}$. The
exact value of $l_{sur}$ depends on $\beta_{1}, \ldots \beta_{k}$, unless
we use the variant of the algorithm described in \ref{dep-delta}.

\begin{thm}\label{g-theorem}

The algorithm presented finds the right element with probability
$1- \delta$ in an expected query complexity of

$$\frac{\log(n)}{I(p_0, \ldots, p_k)} + O(\frac{\log \log(n)
\log(1/\delta)}{I(p_0, \ldots, p_k)})$$


\end{thm}

\section{Quantum Search With a Non Faulty Oracle}\label{sec:quantumSearch}

Farhi et al. presented in \cite{FGGS99} a ``greedy'' algorithm,
which given an array of size $K$ and $t$ queries, attempts to find
the correct element but has some error probability. Their algorithm
actually gives something better. Assume that the elements given to
their algorithm are $y_0,..,y_{K-1}$ and the special element $s$.
Again we are trying to find $i$ which satisfies $y_{i} \ge s \ge
y_{i+1}$ (we use different notation than $x_1, \ldots, x_n$ as we
are going to combine algorithms with $K$ being a constant regardless
of $n$). Their algorithms outputs a quantum register with the
superposition $\Sigma_{j=0}^{K-1} \beta_{j} \ket{(j + i)}$ (with all
indexes taken mod $K$) for fixed $\beta_0, \ldots ,\beta_{K-1}$
which are not a function of $s$. Let $p_j = |\beta_j|^2$, then
measuring this register we obtain the correct value with probability
$p_0$. The exact numbers $p_0, \ldots p_{K-1}$ are determined by the
number of oracle queries $t$. We now use their algorithm (with
proper values for $K$ and $t$) as a subroutine in our generalized
search algorithm with $k = K$.

Using $K = 2^{23}$ and $t = 6$ gives a distribution $Q$ with
$I(p_0, \ldots, p_k) = 18.5625$. This gives us an algorithm which
requires less than $0.32 \log(n)$ oracle questions with $o(1)$
failure probability. This gives theorem \ref{main-quantum}.


%

\section{Quantum Lower Bounds}\label{sec:quantumLowerBounds}

To prove lower bounds we use an oracle similar to the one in
\cite{HMW03} and \cite{BNRW03}. Let $O \prime$ be a quantum oracle,
$O\prime(\ket{xc}) = \ket{x(0 \oplus c)}$ if $x \in L$ and $\ket{x(1
\oplus c)}$ if $x \not \in L$.
To make $O$ noisy, let $O(\ket{xc}) = \cos(\alpha) \ket{x(c \oplus
f(x))} + \sin(\alpha) \ket{x(c \oplus f(x) \oplus 1)}$ where $f(x) =
1$ if and only if $x \in L$, and $\cos(\alpha) = \sqrt{p}$.

%
%





\begin{thm} Any noisy quantum
algorithm requires $\Omega(\log(n) / I(p))$ queries.
\end{thm}

\begin{proof}
Define $\lambda = p - 1/2$. We use notation and techniques of
\cite{HNS02} and assume the reader is familiar with the proof. We
assume that a run of the algorithm consists of $A = (UO)^{T}U
\ket{0}$, where $O$ is an oracle call, $U$ is a unitary and the
algorithm requires $T$ oracle calls. The quantum algorithm is given
an unknown oracle $x$ out of a group $S$, and after the run a
measurement is done and the algorithm guesses which oracle was given
to it. \cite{HNS02} define the state $\ket{\psi_x^j}$ to be the
quantum state after $j$ iterations, when the oracle was $x$. They
define a weight function $W_j = \Sigma_{x,y \in S} \omega(x,y)
\braket{\psi_x^j}{\psi_y^j}$ where $\omega(x,y)$ is an un normalized
distribution on input states. \cite{HNS02} show that if we choose

\[ \omega(x,y) =
   \left\{ \begin{array}{l}
       \frac{1}{h(y) - h(x)} \qquad if \quad 0 \le h(x) < h(y) < n \\
       0 \qquad \qquad \qquad otherwise \end{array} \right. \]

\noindent where $h(x)$ is the hamming weight of x then $W_0 = n
H_n - n$ and $W_T = \delta \prime W_0$, where $\delta \prime =  2
\sqrt{\delta (1- \delta)} $, $H_i = \Sigma_j \frac{1}{j}$ the i'th
harmonic number, and $\delta$ is the probability for the algorithm
to succeed.

To finish the argument, we need to bound the difference between
$W_j$ and $W_{j+1}$ and thus gain a bound on $T$. Define $P_i =
\Sigma_{z \ge 0} \braket{z;i}{z;i}$ the projection operator.  We
deviate a little bit from their article now, and devise a better
bound assuming that the quantum oracle is noisy. \cite{HNS02} use
the fact that $| \braket{\psi_x^j}{\psi_y^j} -
\braket{\psi_x^{j+1}}{\psi_y^{j+1}}| \le 2 \Sigma_{i, x_i \neq
y_i} ||P_i \ket{ \psi_x^j}|| \cdot ||P_i \ket{ \psi_x^j}|| $.

But when the oracle is noisy, we actually have $|
\braket{\psi_x^j}{\psi_y^j} - \braket{\psi_x^{j+1}}{\psi_y^{j+1}}|
\le 2 \Sigma_{i, x_i \neq y_i} ||P_i \ket{ \psi_x^j}|| \cdot ||P_i
\ket{ \psi_x^j}|| \cdot (1 - \sqrt{1 - 4 \lambda^2})$, which is
very close to multiplying with $1 / I(p)$. The proof in
\cite{HNS02} continues by proving an upper bound of $\pi n$ using
this sums. plugging this estimation in their proof gives us a
factor of $(1 - \sqrt{1 - 4 \lambda^2})$. As the maximal expected
weight loss is $\pi n / I(p)$, it would require at least
$\Omega(\log(n) / I(p))$ queries for a quantum algorithm.
\end{proof}



Using our techniques enables us to give a better lower bound for
the number if queries $t$ a quantum noiseless algorithm needs to
the find the right element out of $k$ (note we search $k$ instead
of $n$ elements) with probability $\ge 1 - \delta$. \cite{HNS02}
gave a lower bound of $t \ge ( 1 - 2 \sqrt{ \delta (1 - \delta)})
\frac{1}{\pi} (H_k - 1)$, applicable only for $\delta < 1/2$.

\begin{thm}
Any quantum algorithm which finds the right element with
probability greater than $1 - \delta$ requires $t \ge
 \frac{\ln(2)}{\pi}((1 - \delta) \log(k)) - O(\delta)$  queries.
\end{thm}

\begin{thm}
Any quantum algorithm which finds the right element with probability
greater than $1 - \delta$ requires $t \ge
 \frac{\ln(2)}{\pi}((1 - \delta) \log(k)) - O(\delta)$  queries.
\end{thm}
\begin{proof}

Assume we have such an algorithm.
Plug it as subroutine in \ref{gen-noisy-search}, using $p_0 = 1 -
\delta$, and $p_j = \delta / (k - 1)$ for $j \neq 0$. This would
give $I(p_0, \ldots, p_k) = \log(k) + (1 - \delta) \log (1 - \delta)
+ \delta \log (\delta / (k - 1))$, and an information gain rate of
$I(p_0, \ldots, p_k)/t$ bits of information per query. However, we
know from \cite{HNS02} that any perfect quantum search algorithm for
an ordered list needs at least $\frac{1}{\pi} \ln (n)$ queries. This
means that the average information gain per query can be at most
$\pi / \ln(2)$ bits per query. This means that $\frac{1}{t} (\log(k)
+ (1 - \delta) \log (1 - \delta) + \delta \log (\delta / (k - 1)))
\le \frac{\pi}{\ln(2)}$

And the number of queries $t$ is at least

$ t \ge \frac{\ln(2)}{\pi} (\log(k) + (1 - \delta) \log (1 - \delta)
+ \delta \log (\delta / (k - 1))) \ge \frac{\ln(2)}{\pi}((1 -
\delta) \log(k) - I(\delta) - 1) \approx \frac{\ln(2)}{\pi}((1 -
\delta) \log(k)) - O(\delta) $ \end{proof}

This lower bound improves the previously known lower bound, and also
has a meaning for relatively high error probability $\delta \le
(k-1)/k$, unlike the lower bound of \cite{HNS02} which has a meaning
only for $\delta < 1/2$.

\appendix

%
%
%
%
%

\section{A Review of the Greedy Algorithm}\label{sec:Quantum Greedy Algorithm}

In this appendix we give a short presentation of the quantum
algorithm of \cite{FGGS99}, which is being thoroughly used in our
paper. Farhi et al. look at a problem in the orale model which is
congruent to searching an element in a list. They define $N$ oracles

\[ f_{j}(x) =
   \left\{ \begin{array}{l}
       -1, \mbox{\ x $<$ j} \\ +1, \mbox{ x $\ge$ j}  \end{array} \right. \]

\noindent for $j = 0, \ldots N-1$. The goal of the algorithm is
given access to an oracle which calculates $f_j(x)$ for unknown $j$,
ind $j$. A query to the oracle consists of calculating $f_{j}(x)$ or
some $x$. They continue by defining

\[ F_{j}(x) =
   \left\{ \begin{array}{cr}
       f_j(x), & \mbox{\ $0 \le x \le N-1$} \\ - f_{j}(x-N), & \mbox{ $N \le x \le 2N-1$}  \end{array} \right. \]

\noindent which is important because $F_{j+1}(x) = F_j(x-1)$ where
we identify $-1$ with $2N - 1$. They also define $G_{j}\ket{x} =
F_j(x) \ket{x}$ and $T\ket{x} = \ket{x + 1}$. This means that their
algorithm can be described as

$$ V_k G_j V_{k-1} \ldots V_1 G_j V_0 \ket{0}$$

\noindent Followed by a projective measurement which decides the
result. Noticing that $T^j G_j T^{-j} = G_0$, Farhi et al found a
base which they denote $\ket{0+}, \ldots, \ket{N - 1+}, \ket{0-},
\ldots \ket{N-1 -}$ such that $T^j \ket{0 \pm} = \ket{j \pm}$, and
when the measurement results in $j \pm$, the algorithm outputs that
he oracle is $j$\footnote{Actually the result should be $\ket{j+}$
if k is even and $\ket{j-}$ if k is odd. We ignore this point as it
is not necessary for the understanding of the algorithm.}.

Demanding that $V_l = T V_{l-1} T^{-1}$, it is possible to calculate
the success probability of any given algorithm, by looking at the
inner product $\braket{V_k G_0 V_{k-1} \ldots V_1 G_0 V_0 \ket{0}}{0
\pm}$. For any given state $\ket{\psi}$, it is possible to calculate
which $V$ will maximize $\braket{V G_0 \psi}{0 \pm}$. Farhi et al
define the greedy algorithm recursively starting from $V_0$, such
that each $V_l$ is chosen to maximize the overlap of $\ket{V_{l-1}
G_0, \ldots V_1 G_0 V_0}$ with $\ket{0 \pm}$. Farhi et al. could not
find an asymptotical analysis of this algorithm, and as it has a
probability to err they decided to use another algorithm as a
subroutine for their search algorithm. We calculated the ``greedy''
algorithm for various parameters, and looked also at the overlap
$\braket{V_{l-1} G_0, \ldots V_1 G_0 V_0}{j \pm}$ for $j \neq 0$.
Differences in overlaps with different $j$ values enabled us to get
the error probability distribution we used before as subroutines in
our classical search algorithm.




\begin{thebibliography}{99}

\small{

\bibitem [AD91]{AD91}J. A. Aslam, A. Dhagat ``Searching in the presence
of linearly bounded errors,''  {\it STOC'91}, 486-493, 1991.

\bibitem[Amb99]{Amb99} A. Ambainis, ``A better lower bound for quantum
algorihtms searching an ordered list,'' {\it FOCS'99}, 352-357 1999.


\bibitem[Asl95]{Asl95} J. A. Aslam, ``Noise Tolerant Algorithms for Learning and Searching,''
{\it PhD thesis, Massachusetts Institute of Technology}, 1995. MIT
technical report MIT/LCS/TR-657.


\bibitem[BBHT98]{BBHT98} M. Boyer, G. Brassard, P. Hoyer, A. Tapp
``Tight bounds on quantum searching,'' {\it  Fortschritte der
Physik}, 46(4-5):493-505, 1998.

\bibitem[BK93]{BK93} R. S. Borgstrom, S. Rao Kosaraju ``Comparison based
search in the presence of errors,'' {\it STOC'93} ,130-136, 1993.

\bibitem[BNRW03]{BNRW03} H. Buhrman, I. Newman, H. Rohrig, R. de Wolf
``Robust Quantum Algorithms and Polynomials,'' {\it
  CoRR quant-ph/0309220}, 2003.

\bibitem[CLP06]{CLP06}  A. M. Childs, A. J. Landahl, P. A. Parrilo ``Improved quantum
algorithms for the ordered search problem via semidefinite
programming,''{\it Phys. Rev. A 75, 032335}, quant-ph/0608161 2007.

\bibitem[DGW92]{DGW92} A. Dhagat, P. Gacs, P. Wincler ``On playing
"twenty questions" with a liar,'' {\it SODA'92},16-22 1992.

\bibitem[FGGS99]{FGGS99} E. Farhi, J. Goldstone, S. Gutmann, M. Sipser
``Invariant quantum algorithms for insertion into an ordered list,''
{\it
  quantph 9901059} January 1999.

\bibitem[FRPU94]{FRPU94} U. Feige, P. Raghavan, D. Peleg, E. Upfal
``Computing with noisy information,'' In {\it
  SIAM J. of Computing,} 23, 1001-1018, 1994.

\bibitem[HEWJB04]{HEWJB04} J. Heinemeyer, H. Eubel, D. Wehmh\"{o}nre, L.
J\"{a}nsch, H. Braun ``Protemic approach to characterize the
supramulecular organization of photosystems in higher plants,'' In
{\it Phytochemistry} 65, 1683-1692, 2004.

\bibitem[HMW03]{HMW03} P. Hoyer, M. Mosca, R. de Wolf: ``Quantum Search
on Bounded-Error Inputs ,'' {\it ICALP}, 291-299 , 2003.

\bibitem[HNS02]{HNS02} P. Hoyer, J. Neerbek, Y. Shi ``Quantum
compexitiees of ordered searching, sorting and element distincness
,'' {\it
  Algorithmica,} 34, 429-448, 2002.


\bibitem[JLB05]{JLB05} M. B. Jacokes, A. J. Landahl, E. Brooks
``An improved quantum algorithm for searching an ordered list,``
{\it preparation} 2005.

\bibitem[KK07]{KK07} R. Karp, R. Kleinberg ``Noisy binary search and applications,'' In {\it SODA'07},
881-890, 2007.

\bibitem[KMRSW80]{KMRSW80} D. J. Kleitman, A. R. Meyer, R. L. Rivest, J.
Spencer, W. Kinklmann ``Coping with errors in binary search
procedures,'' {\it
  Journal of Computer and System sciences} 20, 396-404,  1980.

\bibitem[Mut96]{Mut96} S. Muthukrishnan ``On optimal strategies for
searching in the presence of errors,'' {\it SODA'96},680-689 1996.

\bibitem[Orr96]{Orr96} Genevieve B. Orr, ``Removing Noise in On-Line Search using Adaptive Batch Sizes,''  {\it
  Nips 1996},1996.

\bibitem[Ped99]{Ped99} A. Pedrotti ``Searching with a constant rate of malicious lies,'' In {\it
  Internat. Conf. Fun with Algorithms}
,137-147 June 1999.

\bibitem[Pel89]{Pel89} A. Pelc ``Searching with known error probability,''
In {\it Theoretical Computer Science}, 63,185-202, 1989.

\bibitem[Pel02]{Pel02} A. Pelc ``Searching games with errors: fifty years of
coping with liars,'' In {\it Theoretical Computer Science},
270,71-109, 2002.

\bibitem[SCJ94]{SCJ94} H. Sch\"{a}gger, W. Cramer, G. von Jagow
``Analysis of Molecular Masses and Oligomeric States of Protein
Complexes by Blue Native Electrophoresis and Isolation of Membrane
Ptotein Complexes by Two-Dimensional Native Electrophoresis,'' In
{\it Analytical Biochemiostry} 217,220-230, 1994.


}
\end{thebibliography}
\end{document}